\documentclass[letterpaper,10pt]{article} 
\usepackage{opticameet3} %% use version 3 for proper copyright statement

\newcommand\authormark[1]{\textsuperscript{#1}}

\usepackage{amsmath,amssymb}
\usepackage[colorlinks=true,bookmarks=false,citecolor=blue,urlcolor=blue]{hyperref} %pdflatex
\usepackage[mathscr]{euscript}

\begin{document}

\title{On the Bit Error Rate Fluctuation Induced by Multipath Interference in the Coherent Regime for Intra Data Center Applications}

% \author{Author name(s)}
% \address{Author affiliation and full address}
% \email{e-mail address}

\author{Wing-Chau Ng\authormark{1} and Scott Yam\authormark{2}}

\address{\authormark{1}Carleton University, 1125 Colonel By Dr, Ottawa, ON K1S 5B6, Canada\\
\authormark{2}Queen's University, 19 Union Street, Kingston, ON, K7L 3N6 Canada}

\email{\authormark{*}scott.yam@queensu.ca} %% email address is required

%% Do not add a copyright statement. Optica will add it.

\begin{abstract}
We theoretically explain how multipath interference resizes PAM-4 constellation in the coherent regime and thus increases bit error rate fluctuation in intra data centers for the first time. 
\end{abstract}

\section{Introduction}
Recent huge demand on intra data center connectivity for scale-up and scale-out have eventually driven the massive production of 1.6 Tbps (200 Gbps/lane, 2${\times}$DR4/FR4) PAM4 pluggable modules by the end of 2025. In addition to limited analog and optical device bandwidth, optical amplitude-dependent impairments such as laser relative intensity noise (RIN), four-wave mixing (FWM), and multipath interference (MPI) further reduce the system margin \cite{Google2025}. \cite{MetaMPI} first analytically and experimentally analysed the MPI-induced performance degradation for the decorrelation regime (where the optical path-length difference $L$ between the signal and its reflections is much larger than the laser coherence length $L_c$). A recent work \cite{ecoc2025} experimentally discovered an anomalously high bit error rate (BER) randomly occurred in the coherent regime ($L < L_c$), showing its impact even for links shorter than 100 meters. However, the root cause still remains unknown.  

This works aims to provide a theoretical analysis on why the MPI may get worse in the coherent regime compared to the decorrelation regime. Namely, the random phase offset introduced by laser frequency drift and optical path length difference may lead to constellation dilation and contraction. This provides a convincing theoretical support for the experimental findings of the BER fluctuation in \cite{ecoc2025}.   
\section{Formulation}
Considering a single-reflection case, the instantaneous optical signal power $P_{sig}(t)$, proportional to PAM4 symbol data $d(t) \in \mathscr{D} =\{-1.5, -0.5, 0.5, 1.5\}$ shifted by the positive bias $V_b$, at carrier frequency $\omega_o$ with laser phase noise $\theta(t)$, is reflected by uncleaned or improperly installed fiber connectors. The reflection (or called the interferer) travels over an extra round-trip distance $L$ to reach the receiver. Compared to long-haul coherent applications, the laser sources for data center applications usually have a much higher linewidth (few to tens of MHz) without sophisticated frequency control. Thus, the signal carrier frequency experiences a shift of $\Delta \omega$ compared to the interferer's after the round-trip time $\tau = nL/c$. Representing the signal phasor as $e^{j\omega_o t+j\theta(t)}$ while the interferer phasor as $e^{j(\omega_o+\Delta \omega)(t-\tau)+j\theta(t-\tau)}$, the received signal becomes \cite{MetaMPI, ecoc2025, wcNg2025}
\begin{equation}\label{eq:rxSignal}
    \begin{split}  
    y(t)  = d(t) + V_b +  \overbrace{2 \rho \sqrt{ V_{b} + d(t)} \sqrt{V_{b} +d(t-\tau)} \underbrace{\cos{[\Phi +\theta(t)-\theta(t-\tau)]}}_{  B(\Phi, t)}}^{\Delta V_{b, MPI}} + n(t).
    \end{split}
\end{equation}
% Eq.~(\ref{eq:rxSignal})
where $\rho^2$ is the MPI power ratio, $\Phi = \omega_o\tau - \Delta\omega t + \Delta\omega\tau $ is the extra phase offset introduced by laser frequency instability happening over $\tau$ as well as the path-length difference. The MPI-induced bias fluctuation $\Delta V_{b, MPI}$ is composed of self amplitude modulation by the signal itself $\sqrt{V_{b} +d(t)}$, cross amplitude modulation by its time-shifted copy $\sqrt{V_{b} +d(t-\tau)}$ and phase-noise dynamics $B$. In practice, $d(t-\tau)$ is unknown to receivers because of the unknown $L$, and thus acts as a high-frequency interference that no ASIC-rate DSP algorithms can track accurately \cite{wcNg2025}. What DSP attempts to compensate for is the slowly-varying (relative to data rate) bias fluctuation shaped by the envelop $B(\Phi, t)$. In the following, we will explain in detail the impact of $\Phi$ on BER fluctuation for $L < L_c$. 

  \begin{figure}[htbp]
  \centering
  \includegraphics[width=16.3cm]{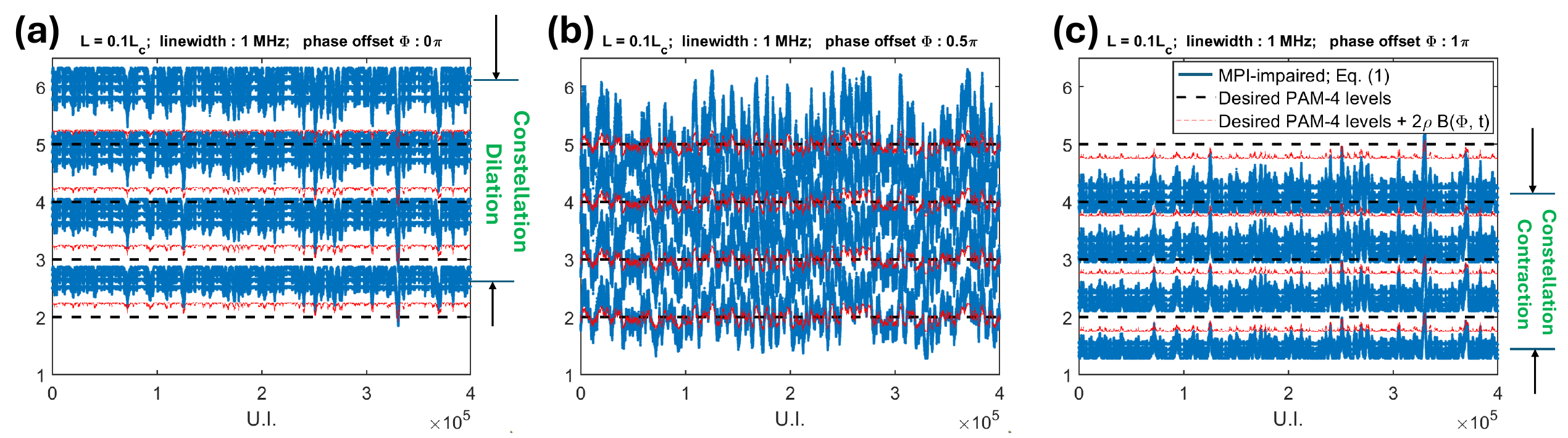}
\caption{ Impact of phase offset on Euclidean distance (a) $\Phi = 0\pi$ (b) $\Phi = 0.5\pi$  (c) $\Phi = 1\pi$ for $L = 0.1L_c$. U.I.: unit interval (symbol duration).  }
\label{fig:waveformExplain}
\end{figure}
\subsection{Phase-offset induced constellation dilation and contraction}
For $L < L_c$, the extra phase offset between the signal and its reflection cannot be overlooked. For example, a round-trip length of 5 meters with  5 MHz laser drift results in $\Delta\omega\tau$ of around ${0.5}\pi$, not to mention the possible extra phase offset from $\omega_o\tau - \Delta\omega t$. Fig. \ref{fig:waveformExplain} shows the MPI-impaired waveforms for three different values of $\Phi$. To help understanding, we provide new reference curves in red, representing the distorted but still equally-spaced PAM-4 reference levels being modulated by phase noise dynamic $B(\Phi,t)$, which follows \textcolor{red}{$\mathscr{D}+V_b+2\rho B(\Phi, t)$}. The self amplitude modulation $\sqrt{V_b+d(t)}$ and phase offset $\Phi$ in Eq.~(\ref{eq:rxSignal}) dilate or contract the constellation when $\Phi$ is in the neighborhood of zero or to $\pi$, respectively, leading to different Euclidean distances. To illustrate mathematically, for instance, assuming that the cross phase modulation $\sqrt{V_b+d(t-\tau)}$ gets averaged out and become $\sqrt{V_b}$\cite{wcNg2025}, by in Eq.~(\ref{eq:rxSignal}), transmitted symbols "0.5" and "1.5" will be received as \textcolor{red}{$R_2 = 0.5+V_b+2\rho \sqrt{V_b+0.5}\sqrt{V_b} B(\Phi, t)$} and \textcolor{red}{$R_3 = 1.5+V_b+2\rho \sqrt{V_b+1.5} \sqrt{V_b} B(\Phi, t)$} , respectively. Their distance $R_3-R_2$ is roughly $ 1+2\rho \sqrt{V_b}(\sqrt{V_b+1.5}-\sqrt{V_b+0.5})$ near $\Phi = 0$ (constellation dilation), and $ 1-2\rho \sqrt{V_b}(\sqrt{V_b+1.5}-\sqrt{V_b+0.5})$ near $\Phi = \pi$ (constellation contraction), respectively. Q.E.D..

Note that the LMS algorithm can track and roughly but not perfectly equalize the shifted mean of the MPI-impaired reference levels. $\Phi = \pi$ should be the worst case, shown in Fig. \ref{fig:waveformExplain}(c) as its Euclidean distance is the smallest. As $\Phi$ can take any values due to random laser frequency drift, the BER is also supposed to fluctuate randomly over time, being upper- and lower- bounded by the above two extreme cases.   
\begin{figure}[htbp]
  \centering
  \includegraphics[width=16.3cm]{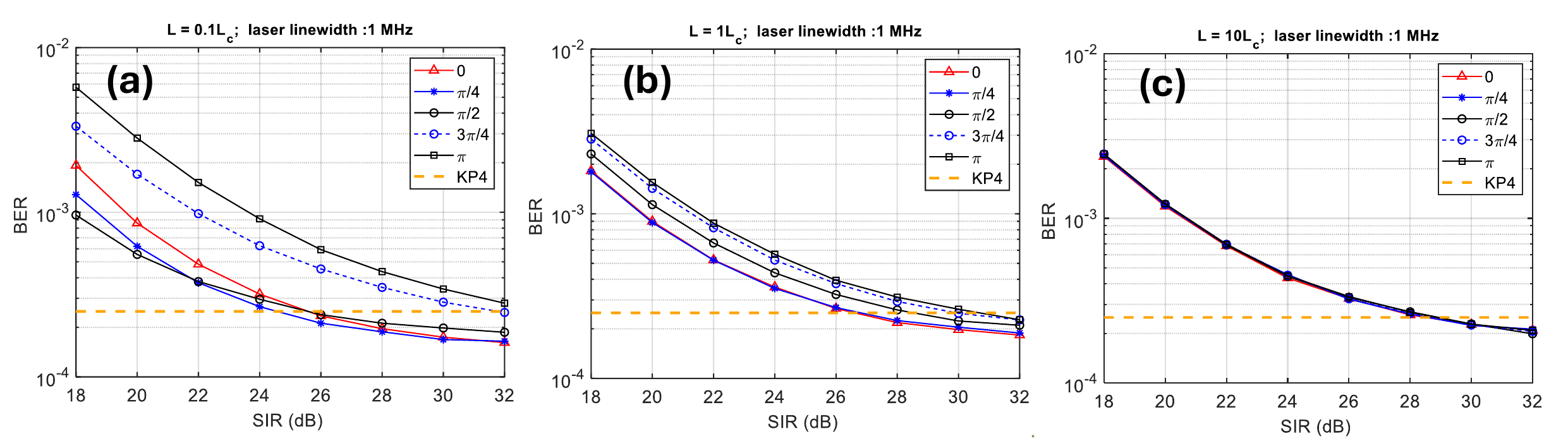}
\caption{Impact of phase offset $\Phi$ on BER over SIR (a) $L = 0.1 L_c$ (b) $L = 1 L_c$ (c) $L = 10 L_c$ at SNR = 18dB.}
\label{fig:fig10MHz}
\end{figure}
\section{Simulation Results}
To analyse the impact of $\Phi$ on BER of 200 Gbps/lane 2$\times$DR4/FR4 modules, 9 millions 106.25GBaud random PAM4 symbols (with gray-coded) were generated at transmitter, with $d(t) = \{-1.5, -0.5, 0.5, 1.5\}$ shifted by the bias $V_b = 3.5$ to yield an extinction ratio of 4dB. Wiener laser phase noise was added in the field domain. A replica of the signal power was attenuated by a specified MPI power ratio $\rho^2$ and delayed by a number of symbols corresponding to $L$. The signal-to-interference ratio is defined as  $SIR = -20\log_{10}\rho$ as if in \cite{ecoc2025}. 
$\Phi$ was added to the laser phase of the reflection to mimic the effect of the phase offset due to laser frequency drift and path-length difference, parameterized with ${0, \pi/4, \pi/2, 3\pi/4, \pi}$ to cover only one principal range of cosine. Intersymbol interference (ISI) is ignored in order to see the pure MPI effect. Additive white gaussian noise (AWGN) was added with respect to the signal power only (excluding MPI power). In the receiver side, a 15-tap feedforward equalizer (FFE) tracked the MPI-induced bias via an additional tap, which was based on the commonely-used least-mean square (LMS) algorithm\cite{biasLMS}. No decision feedback equalizer (DFE) was applied. The BER was then measured given a certain SIR and signal-to-noise ratio (SNR). All the following results were generated at SNR = 18 dB. Fig. \ref{fig:fig10MHz} shows the BER results for three different regimes calculated over 9 million symbols. 

For $L = 0.1 L_c$ (coherent regime), shown in Fig. \ref{fig:fig10MHz}(a), different phase offsets $\Phi$ lead to different BER curves. Interestingly, one can always observe that the black square curve ($\Phi = \pi$) always upper-bounds the BER, while the red curve ($\Phi = 0$) bounds below. This could be explained with the aid of MPI-impaired waveforms. In Fig. \ref{fig:waveformExplain}(a), the phase difference $\theta(t)-\theta(t-\tau)$ is more likely to randomly walk around zero, leading to a more probable positive DC shift. The self amplitude modulation by $\sqrt{V_b + d(t)}$ further pushes the higher levels towards higher amplitudes, leading to a "constellation dilation" as mathematically described previously. The resultant enlarged Euclidean distance then lowers the BER for the case of $\Phi = 0$.   Similarly, in Fig. \ref{fig:waveformExplain}(c),  $\theta(t)-\theta(t-\tau)$ is more likely to randomly walk around $\pi$, leading to a more probable negative DC shift. The self amplitude modulation by $\sqrt{V_b + d(t)}$ results in "constellation contraction". The resultant reduced Euclidean distance then increases the BER for the case of $\Phi = \pi$.  

For $L = L_c$, shown in Fig. \ref{fig:fig10MHz}(b), the phase noises of the signal and its reflection become more decorrelated.  $\theta(t)-\theta(t-\tau)$ is more likely to walk in a larger range of angle, lessening the "initial" impact from $\Phi$. Consequently, the difference between the black and the red curve reduces. 

For $L = 10 L_c$ (decorrelaton regime), shown in Fig. \ref{fig:fig10MHz}(c), the phase noises of the signal and its reflection become decorrelated.  $\theta(t)-\theta(t-\tau)$ randomly walks over any angles, and thus the initial phase offset $\Phi$ has no more impact. All $\Phi$'s yield the same BER result. 

We would here emphasize that Fig. \ref{fig:fig10MHz}(a)-(c) also shows a similar trend compared to the unexplained experimental results shown in Fig. 3 in \cite{ecoc2025}, where the "initial" phase $
\Phi$ was ignored.

\begin{figure}[htbp]
  \centering
  \includegraphics[width=16.3cm]{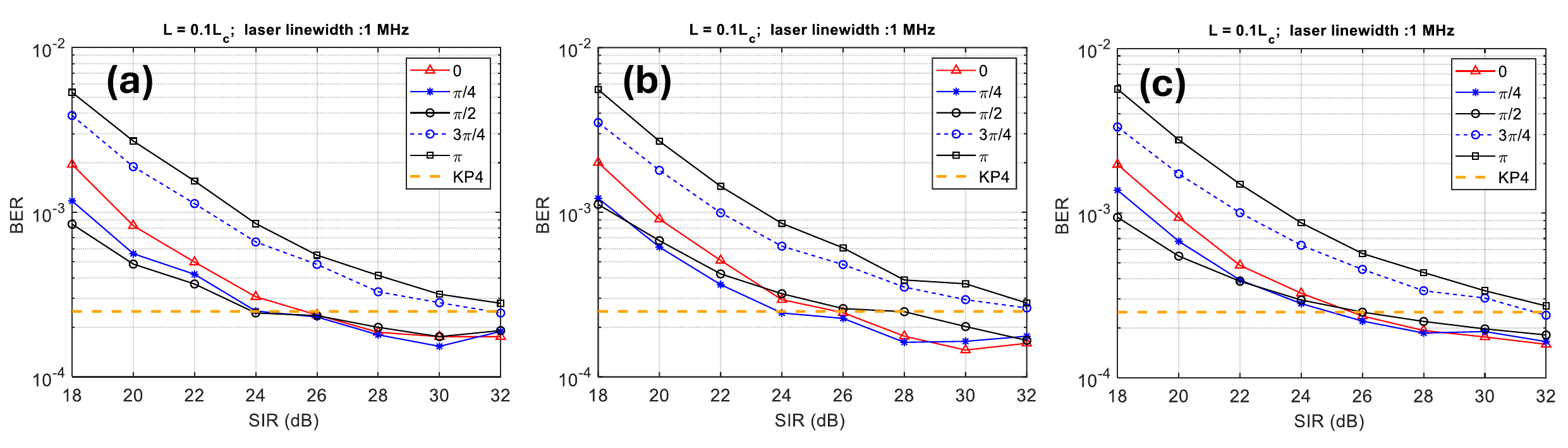}
\caption{ BER results for different number of symbols for simulation: (a) 230,000 symbols (b) 450,000 symbols symbols  (c) 900,000 symbols  }
\label{fig:coherent_numSym}
\end{figure}
As a sanity check, to make sure that the observation time duration can a sufficiently large statistics of phase noise dynamics, we have also repeated the above simulations over different number of symbols, (a) 230,000, (b) 450,000 and (c) 900,000 for simulation as shown in Fig. \ref{fig:coherent_numSym}. Using 230,000 symbols is already enough to catch the trend of the BER fluctuation in the coherent regime, while using 900,000 symbols would yield less statistical variations. 

\section{Conclusion}
In this contribution, we theoretically explain how the random phase offset (caused by laser frequency drift and the path-length difference, but not caused by Wiener phase noise) between the signal and its reflection(s) leads to constellation dilation and contraction, increasing BER fluctuation in the coherent regime ($<$100 meters) in intra data centers. For example, the phase offset caused by the laser frequency drift could randomly cover all possible angles as the DR4/FR4 laser sources do not have sophisticated frequency. As the round-trip length increases, the phase noises of the signal and its reflections become more decorrelated, whose effect dominates the phase offset, thus averaging out the BER fluctuation.

\section{Acknowledgement}
 The authors express their gratitude to Dr. S. Oettinghaus for generously sharing his Matlab code to reproduce Fig. 1(b) of his recent work \cite{ecoc2025} as well as fruitful discussion.

\end{document}